\documentstyle [12pt]{article}
\begin{document}
\begin{titlepage}
\begin{flushright}
ILL-(TH)-92-\# 15\\
JUNE 1992\\
ROM2F-92-46
\vskip 1.5truecm
\end{flushright}
\begin{center}
{\LARGE Lattice QCD Spectroscopy with an Improved Wilson Fermion Action }\\
\vskip 1.0truecm
{\large Maria-Paola Lombardo}
\footnote{ on leave from Infn, Sezione di Pisa, Italy}\\
\vskip 0.2truecm
{\normalsize  Department of Physics,}\\
{\normalsize  University of Illinois at Urbana-Champaign,}\\
{\normalsize  1110 West Green Street, Urbana, IL 61801, U.S.A.}\\
\vskip 1.0truecm
{\large Giorgio Parisi and Anastassios Vladikas}\\
\vskip 0.2truecm
{\normalsize  Dipartimento di Fisica,}\\
{\normalsize  Universit\`a di Roma {\it Tor Vergata}}\\
{\normalsize  Via E.Carnevale, 00173 Roma, Italy} \\
{\small  and} \\
{\normalsize  Infn, Sezione di Roma {\it Tor Vergata}} \\
\vskip 1.0truecm
\end{center}
\end{titlepage}
\vfill
\newpage

\begin{abstract}
We study the hadronic spectrum in quenched lattice QCD using the improved
Wilson fermion action proposed in~ \cite{HW}
at $\beta= 5.7$ and $\beta =6.0$.
We find a systematic reduction
of the finite spacing effects compared to the results
obtained by using the standard Wilson action.
\end {abstract}
\vfill
\newpage

\section{Introduction}
{}~~~~Most of the  recent results about quenched QCD
spectroscopy~\cite{DEGRAND_FSU}
exhibit the scaling behaviour expected near the continuum
limit~\cite{W63}.
However, deviations are still observed in the phenomenologically
relevant heavy quark mass region,
and in these cases progress
using the standard Wilson action seems to be slow.
Moreover, an important self-consistency check of lattice computations
is the agreement between results coming from different discretizations.

For the above two reasons, we decided to study lattice spectroscopy
by using an improved version of the  Wilson fermion action \cite{HW},
inspired by
Symanzik's perturbative procedure of eliminating finite-$a$
corrections \cite{SY}.
Other  proposals along the same line are the ones of
ref.~\cite{SW} and ref.~\cite{IT}, which,
by combining the improved action with a suitable modification
of the fermion operators,
eliminates all terms of order $a$ in the  hadronic matrix elements.
Some results obtained by using the actions proposed in
refs.~\cite{SW},~\cite{IT} are already available and
we remand to the references ~\cite{IR},~\cite{UK},~\cite{FNAL}
for their discussion.

There are  two  points to be considered, in relation to
an improvement program based on perturbative arguments.
The first one is the real
source of the scaling violations, which can be completely non-perturbative,
the other one is the practical applicability. Indeed, the success of
an improved action
depends on the interplay between the  reduction of $O(a)$ effects,
and
the possibly induced new systematics, and it is not guaranteed a priori.
To understand the real effectiveness of the action ~\cite{HW}
we  study in this paper $\beta = 5.7$ and $\beta = 6.0$. From
the results for the standard Wilson case we know that the finite
spacing effects are strong at $\beta= 5.7$, while $\beta= 6.0$ seems
to be the onset of  scaling for many relevant quantities.
We thus use the data at $\beta = 5.7$ to demonstrate the suppression
of the most evident scaling violations which occur using the Wilson action
at the same value of the coupling, and the ones at $\beta = 6.0$ to
explore the possibility to actually improve the current Wilson results.

We introduce the action we use, and  summarize its basic properties
in the next section.  Then, we describe the numerical simulation
and the analysis procedure. Finally we discuss the results for the hadron
spectrum and the meson decay constants.  In the following
we  keep as a reference the results obtained with
the standard Wilson action at $\beta = 5.7, \beta = 6.0,
\beta = 6.3$ of ~\cite{W63},~\cite{W57},~\cite{W60}. Our conclusions
are based on this comparative analysis.

\section{The action}

{}~~~~We use the improved  fermion action proposed by Hamber and Wu~ \cite{HW}.
A next nearest neighbor term is added to the Wilson fermion action:
\begin{equation}
S = S_G + S_W + S_{II}
\end{equation}
$S_G$ and $S_W$ are the usual pure gauge and Wilson fermionic term, while
the new term reads:
\begin{equation}
S_{II} =\sum \bar\psi_n(C-D\gamma_\mu)U_{n,\mu}U_{n+\mu,\mu}\psi_{n+2\mu}
+ h.c.
\end{equation}
The choice $C = - 1/4kr$ cancels at tree level the term $O(p^2)$
in the inverse
propagator $S_F^{-1}(p)$, $D$ is free. We choose $D = -1/8kr$ which cancels
also the terms $O(p^3)$ in the inverse fermion propagator, and set $r = 1$.
In this way the fermionic part of the action coincides with the
Eguchi-Kawamoto one.

We refer to the original references~\cite{HW}, and to ref.~\cite{IT} for
a detailed discussion. It is however interesting pointing out here
that this action exhibits positivity violations
stronger that the ones of the standard Wilson case. The zero-momentum
free quark propagator $S_F(p_0, \vec{0})$ has four poles for any $k$ value,
two of them complex, the other two turning from complex to real for
$k > \simeq 0.14$.( The critical value $k_c$ ,
defined by $m_q(k_c)=0$, $m_q$ being the quark mass, is $1/6$.)
So,
the free propagator, computed  by Fourier transforming $S_F(p)$
to the coordinate space, deviates from a simple exponential behaviour.
It shows a clear ripple at small $k$'s (i.e., in the region of
four complex poles), which is more evident in small
lattices. The effect progressively disappears by increasing the
size of the lattice, and approaching $k_c$, suggesting that
no problem  arises in the continuum limit. Indeed,
in the  continuum limit $m_{PHYS}a = m_{LATT} \to 0$,
while $k \to k_c$.
So, in order to define properly the $a\to 0 $ limit
only one real pole (the one corresponding to the
physical mass) should be zero as $k \to k_c$,
while the other real pole, and the real parts of the complex ones,
should be different from zero.
It is  trivial to show that this is the case.
Turning to the  amplitudes, no problem should arise with the
ones involving  physical status: violations of
positivity only affects the residuals of non-physical poles.

The above informal discussion, which leads to the conclusion
that the spurious poles are harmless, concerns the free case.
However, the results for
the standard Wilson action suggest that what has been found in
the free case is general: for the Wilson action,
the free propagator has a complex pole for $K> 1/6$, and a real
one otherwise, and it has been shown~\cite{WOS}
that the same pattern of violation of
positivity is maintained also in the interacting case.
In conclusion, a more detailed investigation  of the positivity
properties of the improved actions would be welcome, but
at the moment we
have no reason to think that their continuum limit is suspect.

Anyway --- even if this may seem
rather paradoxical --- an action supposed  to reduce
finite size effects potentially leads to a peculiar finite
spacing systematics, whose actual impact is not possible to
estimate a priori.  Happily enough, positivity
violations affected the results
only at $\beta = 5.7$,  for the $L=1$ mesons.

\section{The Numerical Simulation and the Data Analysis }

{}~~~~We used the Wilson pure-gauge action, and we have thermalized the
gauge fields at $\beta = 5.7$ and $\beta = 6.0$. We used a standard
Metropolis-5 hits algorithm to generate the background
gauge field configurations. The onset of the scaling region for
the standard Wilson action seems to be around $\beta = 6.0$
while the results at $\beta = 5.7$ are definitively affected
by finite spacing effects. Thus, the values we choose are good
to assess the effectiveness of the method.

We explored the heavy quark region, which is appropriate
to test this action. We choose four equispaced  K, ranging from
0.186 to  0.198 at $\beta = 5.7 $, and from 0.176 to 0.188
at $\beta= 6.0$.
This gives the ratio $\pi^2/\rho^2$ between 0.4 and 0.8 at $\beta = 5.7$
and  between 0.6 and  0.9 at $\beta = 6.0$.

One configuration every
800  iterations was sampled for the propagator inversion.
At $\beta= 5.7$ we have computed  (140 $\times$ 4 quark masses) propagators
on  a $24\times 12^3$ lattice,  at $\beta = 6.0$ we have (160 $\times$
4) propagators
on a $32 \times 12^3$ lattice. We also collected
$(180  \times 3)$ propagators on a $18\times9^3$ lattice (at the three
bigger masses) to
monitor the finite size effects.

All the measures were performed according to the methods introduced
in ~\cite{W57} and reviewed in ~\cite{ENZO_FNAL}, which we briefly summarize
in the following.
We compute the fermionic propagators
by a preconditioned minimal residue algorithm, as modified
for parallel updating. We apply the incomplete LU
decomposition as a preconditioner, making use in this step of
the standard Wilson operator.
In practice the original equation
for the improved propagator
\begin{equation}
D  X = B
\end{equation}
is replaced by
\begin{equation}
U^{-1}L^{-1}D  X = U^{-1}L^{-1} B
\end{equation}
where L and U are defined in the usual way
\cite{W57} in term of the Wilson fermionic operator.
We have indeed found also with the improved action a considerable
speed-up in the convergence rate by using this preconditioning.

We  use smearing~\cite{ENZO_FNAL}~\cite{W57}:
 after the necessary gauge fixing we solve the
Dirac equation with an extended source (a $3^3$ spatial cube).
The propagators  computed in this way can be smeared on the final point too.
{}From the  quark propagators we form the contractions needed
to evaluate the hadronic Green functions $G_0$
(smeared only at the origin) and $G_x$ (at the origin and at the final point).
We preferred the second ones ($G_{x}$, smeared at the origin and in x) since
$G_{0}$ sometimes tends to amplify the problems related to the lack
of positivity of this action, but the difference is small and  on the
overall the results obtained from $G_{0}$ and $G_{x}$ are fully consistent.
A joint fit of $G_0$ and $G_x$ does not further reduce the errors, which are
always computed by a standard jacknife analysis.

The results for the hadron masses
quoted in Table~\ref{57M} and in
Table ~\ref{60M} are from a single mass fit (for $t > 5$)
of the Green functions
smeared both at the origin and at the final points.
We systematically  controlled the stability of the results
by performing two particle fits as well (in this case  we  included points
starting form $t = \simeq 3$), and we cross checked with  the local fits.
As an example of the overall quality of the data  we show in
Fig.~1 the effective masses of the proton at $\beta = 5.7$
 and in Figs.~2 the fits at $\beta = 6.0$ for the proton and the pion.

\begin{table}
 \begin {tabular} {||l||l l l l ||} \hline
 &\multicolumn{4}{||c||}{\em k  } \\ \hline
 Particle & 0.186 & 0.190 & 0.194 & 0.198 \\
  \hline
      $\pi$    &1.0267( 28)&0.8661( 29)&0.6975( 31)&0.5069( 40)\\
 $\tilde \pi$
\footnote{
As usual, we define the $\tilde\pi$ and $\tilde\rho$
operators by inserting a $\gamma_0$ in the $\pi$ and in the $\rho$.}
  &1.0239( 39)&0.8624( 42)&0.6925( 50)&0.4975( 73)\\
   $  \rho $   &1.1432( 40)&1.0109( 44)&0.8851( 59)&0.7676(132)\\
 $\tilde \rho$ &1.1451( 44)&1.0141( 51)&0.8905( 73)&0.7811(161)\\
  $      p $   &1.7972(108)&1.5811(116)&1.3616(141)&1.1186(277)\\
  $ \Delta$    &1.8646(138)&1.6665(150)&1.4761(198)&1.2990(557)
 \\
 \hline
 \end{tabular}
\protect\caption{Hadron masses as a function of $k$ at $\beta = 5.7$
on the $24\times 12^3$ lattice}
\protect\label{57M}
\end{table}

Using the data on the $9^3\times 18$ lattice we checked that finite
size effects are small. The big symbols in Fig.~1
which perfectly circle the results obtained on the $12^3\times 24$
lattice are from the smaller one. The full set of results on
the $9^3\times 18$ lattice is shown in Table~\ref{57MS}.
(We note that the plateau in the effective masses in this lattice
is not as extended as in the bigger one, so in this case
we had to rely on two particle fits.)
The results reported in Table~\ref{57MS} agree with the ones of
Table~\ref{57M}, thus demonstrating the good
control we have over finite size effects.
We are confident that, since the time extents in physical units
of the lattices  at $\beta = 5.7$ and $\beta= 6.0$ are roughly the
same,  the masses evaluated at $\beta = 6.0$ do not suffer from finite
size effects as well. (However,
the number of points in the space directions is
the same for the two lattices, so some problem at least with the smallest
mass at $\beta = 6.0$ cannot be excluded.)

\begin{table}
 \begin {tabular} {||l|l l l  ||} \hline
 &\multicolumn{3}{||c||}{\em k  } \\ \hline
 Particle & 0.186 & 0.190 & 0.194 \\
  \hline
     $\pi$    &1.0199( 30)&0.8638( 31)&0.6970( 37) \\
$ \tilde \pi$  &1.0207( 40)&0.8623( 46)&0.6714( 59)\\
$     \rho $   &1.1326( 43)&1.0073( 47)&0.8863( 57)\\
$ \tilde \rho$ &1.1327( 48)&1.0052( 56)&0.8814( 75)\\
$        p $   &1.6807(182)&1.5465(126)&1.3505(100)\\
$   \Delta $   &1.6541(294)&1.5798(219)&1.4707(212)
 \\
 \hline
 \end{tabular}
\protect\caption{Hadron masses as a function of $k$ at $\beta = 5.7$
on the $18\times 9^3$ lattice}
\protect\label{57MS}
\end{table}

At $\beta = 5.7$ we do not quote the results for the $a1$ and $b1$
particles. These mesons
deserve some separate comments, since their behaviour turned out
to be quite peculiar.  Their Green functions  can even change
sign, thus indicating violations of positivity (
amplitudes of different signs in different channels,
or  manifestations of the spurious complex poles).
As a consequence, the effective masses have a wiggle
which becomes less and less apparent as $k\to k_c$ or $a \to 0$,
in agreement with the qualitative discussion of Sect.2 above.
In fact, the results at $\beta= 6.0$
for the fits of the
$L=1$ mesons are reasonable. (Of course in this case we do not
have any hint about residual finite size effects, and it is clear that
having amplitudes of comparable magnitude and opposite signs is the worst
possible situation from this point of view.)  As a general comment, the
orbitally excited hadrons are difficult to treat, also with the ordinary
Wilson action, and their signal is
intrinsically noisy: a recent discussion, together with new results, can
be found in ref.~\cite{DGHE}.

We conclude the description of the analysis procedure by discussing the
computations of the amplitudes of the Green functions in the fundamental
channels,
which are important for the estimates of the decay constants. We
are interested in the local amplitudes, which, as discussed in ref~\cite{W60}
can be recovered from the ratio of $G_{0}^2$ to $G_{X}$ in the
hypothesis of long-distance factorization of the Green functions.
We checked that the amplitude of the fit of the ratio has a smaller
statistical error compared to the ratio of the amplitudes of the fits,
typically by a factor 2. This is understandable, because by fitting the ratio
we get rid of the coherent fluctuations. So the errors we quote are
obtained in this way (the central values are basically
the same). At $\beta = 5.7$ we also computed 24 propagators with a point
source, so we can evaluate directly the local amplitudes which turn out
to be in complete agreement with the ones coming from the fit of the
ratios. The summary
of the results for the amplitudes is given in Tables~\ref{57A} and
{}~\ref{60A}.

\begin{table}
 \begin {tabular} {||l|l l l l ||} \hline
 &\multicolumn{4}{||c||}{\em k  } \\ \hline
 Particle & 0.176 & 0.180 & 0.184 & 0.188 \\
  \hline
    $\pi$     &1.0014( 15)&0.8179( 17)&0.6287( 23)&0.4204( 39)\\
 $\tilde \pi$ &0.9976( 17)&0.8141( 21)&0.6252( 29)&0.4170( 52)\\
   $\rho$     &1.0424( 19)&0.8712( 24)&0.7024( 33)&0.5333( 63)\\
 $\tilde \rho$&1.0408( 20)&0.8691( 25)&0.6990( 36)&0.5238( 76)\\
     $a1$     &1.2794(142)&1.1203(187)&0.9676(285)&0.8150(570)\\
     $b1$     &1.2772(140)&1.1159(177)&0.9553(262)&0.7873(557)\\
      $p$     &1.6356( 43)&1.3734( 55)&1.1126( 80)&0.8458(150)\\
 $\Delta$     &1.6602( 48)&1.4061( 61)&1.1584( 89)&0.9120(183)\\
 \hline
 \end{tabular}
\protect\caption{Hadron masses as a function of $k$ at $\beta = 6.0$
on the $32\times 12^3$ lattice}
\protect\label{60M}
\end{table}

\begin{table}
 \begin {tabular} {||l|l l l l ||} \hline
 &\multicolumn{4}{||c||}{\em k  } \\ \hline
 Particle & 0.186 & 0.190 & 0.194 & 0.198 \\
  \hline
    $\pi ^1$ & 2.140( 57)& 1.950( 53)& 1.834( 56)& 1.876( 78)\\
    $\pi ^2$ & 2.123(100)& 1.904(105)& 1.739(112)& 1.647(141)\\
 $\tilde\pi ^1$ & 0.308( 14)& 0.202( 10)& 0.119(  7)& 0.058(  5)\\
 $\tilde\pi ^2$ & 0.299( 26)& 0.192( 21)& 0.112( 16)& 0.058( 14)\\
 $\rho  ^1$    & 0.977( 38)& 0.779( 33)& 0.611( 33)& 0.484( 53)\\
 $\rho  ^2$    & 0.912( 61)& 0.714( 51)& 0.538( 42)& 0.358( 54)\\
 \hline
 \end{tabular}
\protect\caption{Amplitudes in the fundamental channel  as a function
of $k$ at $\beta = 5.7$ on the $24\times 12^3$ lattice, from the ratios'
fit$^1$  and from local operators$^2$}
\protect\label{57A}
\end{table}

\begin{table}
 \begin {tabular} {||l|l l l l ||} \hline
 &\multicolumn{4}{||c||}{\em k  } \\ \hline
 Particle & 0.176 & 0.180 & 0.184 & 0.188 \\
  \hline
    $\pi$    & 0.675( 15)& 0.540( 14)& 0.427( 13)& 0.342( 17)\\
 $\tilde\pi$ & 0.183(  4)& 0.116(  3)& 0.065(  2)& 0.027(  1)\\
   $\rho$    & 0.340(  8)& 0.242(  7)& 0.160(  6)& 0.092(  6)\\
 \hline
 \end{tabular}
\protect\caption{Amplitudes in the fundamental channel  as a function
of $k$ at $\beta = 6.0$ }
\protect\label{60A}
\end{table}

\section{Hadron masses, decay constants and scaling behaviour}

{}~~~~We begin the discussion of our results with the chiral behaviour.
Since we were particularly concerned with heavy masses,
the chiral extrapolation, especially at $\beta = 6.0$, is somewhat delicate.
For heavy flavors,  the data should follow the predictions
of the potential models for quarkonium, whose curvatures as a function
of the quark mass are different from the ones in the
chiral limit. Consequently, it is difficult to decide when a fit,
even if satisfactory on statistical grounds, correctly describes the
chiral behaviour.
With this warning in mind, we fitted the squared hadron masses
as second order polynomials in the quark masses, defined as
$m_q = 2/3(1/k - 1/k_c)$. This turned out
to be the most effective parametrization, in agreement with previous
experiences with heavy masses.

The first step is the determination of $k_c$ which is obtained by
demanding  $m_\pi(k_c) = 0.$
We get at $\beta = 5.7$
\begin{equation}
m_\pi^2  = 2.687(49)m_q + 2.492(84) m_q^2
\end{equation}
and  $k_c = 0.20333(12)$
At $\beta = 6.0$ we have
\begin{equation}
m_\pi^2 = 2.032(40) m_q + 3.504(46) m_q^2
\end{equation}
and $k_c = 0.19216(12)$

It is interesting to compare with the analytical predictions for
$k_c$ which are  poorly reproduced on the lattice in the standard
Wilson case.
We have a  9 percent deviation at $\beta = 6.0$
and 15 percent at $\beta = 5.7$
from the analytical result $k_c = 1/6(1 + 1/18 g^2 + O(g^4))$ (second
entry of ref. ~\cite{HW}),
to be compared with 12 percent and 17 percent for the
standard Wilson action
at the corresponding $\beta$ values.

\begin{table}
 \begin {tabular} {||l|l l l  ||} \hline
 Particle & $m_0^2$ & $s_1$ & $s_2$ \\
  \hline
   $\pi$          &     0      & 2.687(  49)& 2.492(  84)\\
   $\tilde\pi$    & 0.000(   0)& 2.594(  78)& 2.747( 219)\\
   $\rho$         & 0.387(  38)& 1.984( 278)& 3.356( 517)\\
   $\tilde\rho$   & 0.422(  48)& 1.775( 358)& 3.716( 663)\\
   $p$            & 0.551( 114)& 7.648( 828)& 3.635(1572)\\
   $\Delta$       & 1.150( 282)& 5.377(2224)& 7.322(4247)\\
 \hline
 \end{tabular}
\protect\caption{Coefficients of the extrapolating polynomial for the
hadron masses at $\beta = 5.7$}
\protect\label{57E}
\end{table}

The results for the pion at $\beta = 5.7$ and $\beta = 6.0$ with the
results fits superimposed are shown in Fig.~3. In the same figure we also
show the results for the $\tilde\pi$'s, which turn out to be
in full agreement with the ones for the $\pi$.

Once $k_c$ is determined, we go ahead with fits in other channels.
As already said, they are completely satisfactory on statistical grounds,
and do not deserve any special comments, so
we simply quote in Table~\ref{57E} and~\ref{60E} the coefficients of the
extrapolating polynomials,
\begin{equation}
M^2_{hadron} = M^2_0 + s_1m_q + s_2m_q^2
\end{equation}
for the two $\beta's$.

\begin{table}
 \begin {tabular} {||l|l l l  ||} \hline
 Particle & $m_0^2$ & $s_1$ & $s_2$ \\
\hline
$\pi$          & 0          & 2.032(  40)& 3.504(  46)\\
$\tilde \pi$   & 0.000(   2)& 1.996(  31)& 3.541(  85)\\
$\rho$         & 0.119(  10)& 1.872(  63)& 3.658( 108)\\
$\tilde \rho$  & 0.103(  12)& 1.971(  78)& 3.474( 136)\\
$a1$           & 0.436( 146)& 2.742( 942)& 3.217(1698)\\
$b1$           & 0.362( 136)& 3.180( 864)& 2.520(1548)\\
$p$            & 0.290(  36)& 4.926( 198)& 8.047( 323)\\
$\Delta$       & 0.422(  51)& 4.718( 316)& 8.194( 541)\\
 \hline
 \end{tabular}
\protect\caption{Coefficients of the extrapolating polynomial for the
hadron masses at $\beta = 6.0$.}
\protect\label{60E}
\end{table}

We remark again the agreement between the results for the $\pi$ and
the $\tilde\pi$ particles,
and for the $\rho$ and the $\tilde\rho$ particles.

We note that at $\beta = 5.7$
the result $m_\rho(0)$ for the extrapolated $\rho$
mass lies in between the Kogut-Susskind~ \cite{KS57} result and the
Wilson one, supporting the intuitive picture of an ``effective
spacing" between $a$ and $2a$ for the improved action:
\begin{equation}
m_\rho(0)(W) = 0.185(18) < m_\rho(0)(WI) = 0.387(38)
< m_\rho(0) (KS) = 0.76(5)
\end{equation}
At $\beta= 6.0$ the results for Wilson and Wilson improved
are mutually consistent, while the result for Kogut-Susskind
is getting closer to them (the ratio of the K-S results to the Wilson
one is $1.14(7)$ at $\beta= 6.0$):
\begin{equation}
  m_\rho(0)(W) = 0.111(5) \simeq m_\rho(0)(WI) = 0.119(10)
< m_\rho(0) (KS) = 0.144(27)
\end{equation}
As already noticed in~ \cite{KS60}, $\beta = 6.0$  seems the onset
of the region in which the details of lattice discretization
are forgotten.

The comparison of the results obtained with the Wilson action and
with the improved one
is better done by using adimensional quantities: for this
purpose we
give in Table~\ref{EX} the values of some relevant ratios. For the string
tension at $\beta = 5.7$ we use the result of ref~\cite{SCA1}:
$\sigma a^2 = 0.056(2)$. At $\beta = 6.0$ and $\beta = 6.3$ we get, by
analyzing the same configurations we were using
in ~\cite{W60},~\cite{W63}, $\sigma a^2 = 0.0471(35)$ and
$\sigma a^2 = 0.01704(8)$, respectively~\cite{SCA2}.
(Let us remind that
the experimental values to be compared with are   $p/\rho =  1.22$,
$ p/\sigma =  2.24$,  $ \rho/\sigma =   1.83$, assuming the accepted
estimate for $\sigma$ of 420 MeV.)

It is remarkable the full agreement of the improved
results at $\beta= 5.7$
with the ones obtained with the standard Wilson action at $\beta= 6.3$,
which in turn match the experimental results.
At $\beta = 6.0$ the situation is less clear (the $\rho$ is
the same as the normal Wilson case at the same $\beta$,
the proton is closer to the one at $\beta= 6.3$,
but as previously
discussed we do not have a safe estimate of the systematic errors induced
by the chiral extrapolation at $\beta = 6.0$)

\begin{table}
 \begin {tabular} {||l|l l| l l l ||} \hline
        & 6.0 & 5.7 & 6.3 W & 6.0 W & 5.7 W\\
  \hline
     $p/\rho$& 1.561(117)& 1.193(136)& 1.247( 95)& 1.298( 53)& 1.451( 75) \\
   $p/\sigma$& 2.481(179)& 2.124(220)& 2.220(166)& 1.993(101)& 2.217( 93)\\
 $\rho/\sigma$& 1.590( 89)& 1.780( 88)& 1.780( 64)& 1.535( 66)& 1.528( 49)\\
 \hline
 \end{tabular}
\protect\caption{Relationship between the extrapolated values for
$m_q=0$ of the $\rho$ and proton masses and the string tension}
\protect\label{EX}
\end{table}

In the following we will discuss finite mass data, using occasionally
the extrapolated values for the $\rho$ mass
to set a common scale for the results obtained at different $\beta$
values.

An interesting quantity to look at is the splitting between pseudoscalar
and vector mesons with the
same flavour content. Experimentally, $M^2_V - M^2_{PS}$ is known
to satisfy
$ (\rho^2 - \pi^2) > (K^{\star 2} - K^2) \simeq  (D^{\star 2} - D^2)
\simeq  (B^{\star 2} - B^2) $. The available lattice data fail to
reproduce the approximate plateau exhibited at large quark masses.
Since we have to  reproduce a $\vec\mu_1 \vec\mu_2
\delta(\vec {r_{12}})$ interaction,
the hyperfine splitting is a natural candidate for the improvement.

To show the trend in the splitting, we compute the average derivative
$ <d (m^2_\rho - m^2_\pi) / d m^2_\pi> $ for the various cases of interest.
This is done via the linear fit
\begin{equation}
 (m^2_{\rho} - m^2_{\pi}) = S m^2_\pi + Cm^2_\rho(0)
\end{equation}
(note that both $S$ and $C$ are dimensionless)
The summary of these results is given in Table~\ref{TS}.
$C$ should be consistent with $1$ when both $m_{\pi}^2$ and $m_{\rho}^2$
are linear in $m_q$. In this respect, it is interesting to note that $C$
is roughly consistent with $1$ also when we the quark masses are large.
The slope decreases with $\beta$ both for the standard and
for the improved Wilson
actions. Again, the results at $\beta= 5.7$, improved, are
consistent with the ones at $\beta= 6.3$ , standard Wilson, while the
improved data at $\beta= 6.0$ have definitively the smaller slope, closer
to the  experimental results (however, still inconsistent with them).
Anyway, since the slope is a (slowly varying) function of the quark
mass, a more detailed comparison is done by simply superimposing the results
in units of the extrapolated $\rho$ mass.
This is done in Fig.~5. We show in Fig.5a the improved data
alone, and the collection of results in Fig.5b.

\begin{table}
 \begin {tabular} {||l|l l| l l l ||} \hline
        & 6.0 & 5.7 & 6.3 W & 6.0 W & 5.7 W\\
  \hline
   $S$& -0.027(9)&-0.088(23)& -0.078(11)& -0.173(18)& -0.207(17) \\
   $K$& 0.92(6) &0.88(5)& 0.91(2) &1.01(2) & 0.97(2)\\
  \hline
 \end{tabular}
\protect
\caption{Results of the fits for
$(m^2_\rho - m^2_\pi)  = Sm^2_\pi + Cm^2_\rho(0)$}
\protect\label{TS}
\end{table}

The results for the Wilson action
 at $\beta= 5.7$ are more steeper than the other ones in the
overlapping region of $\pi$ masses
($0.2 \le \simeq m^2_\pi \le 2.0$),
while at $\beta = 6.0$ and
$6.3$ the residual difference in the local slope is very small
(from the plot it is clear that the difference in the  average slopes
comes mostly from the contribution at large masses at $\beta= 6.3$,
and from the one at small masses at $\beta = 6.0$).
Turning to the improved action, the  results at $\beta = 5.7$ closely
follow the ones at
$\beta = 6.3$, Wilson. The slope of the data at $\beta = 6.0$,
which can be compared to the one of $\beta = 6.3$, Wilson up to
$m^2_\pi \simeq 4.$, is clearly more flat,
so closer to the experimental data.

The standard Wilson results seem to have a very slow evolution with $\beta$.
In addition  the results at $\beta= 6.2$ reported in \cite{HQ}
which explore up to $(\pi/\rho (0))^2 \simeq 30$ do not show any reversal
in the decreasing trend , while the UKQCD collaboration
 reported no improvement in the hyperfine
splitting at the same $\beta$ value with the Clover action~\cite {UK}
(this in the small mass range). On the contrary,
the data for the improved action
from $\beta = 5.7$ to $\beta = 6.0$
are still appreciably moving in the right
direction, and it is not unreasonable to expect  good results
at a feasible $\beta$ value (i.e. $6.3 - 6.5$).
Interesting results about the splittings are
the ones obtained by the FNAL group~ \cite{FNAL}, which
use the action~\cite{SW}. They observe that the spin splittings are very
sensitive to the parameter $c$ in the $O(a)$ correction term,
$-i/2c\bar\psi\Sigma_{\mu\nu}F_{\mu\nu}\psi$
which contributes mainly an additional magnetic
interaction to the quarks.

Before turning to the baryons, we discuss the results for the meson
decay constants which are given in Table~\ref{Fpirho}.
As explained in \cite {W60}, we obtain
$f_{\pi}$ from the local amplitudes in the $\tilde\pi$ channel whose
computation has been described in Sect. 3 above. The results are
also plotted in Fig.5a, and again we compare with the Wilson ones
in Fig. 5b.

The results obtained by making use of the  Wilson action at
$\beta = 6.0$ and $\beta= 6.3$ are in moderate disagreement,
while we observe a perfect coincidence between the results  of the improved
action at $\beta= 5.7$ and $\beta= 6.0$ with the ones at $\beta= 6.3$, Wilson.
This suggests that the data at $\beta= 6.3$ Wilson or, equivalently,
$5.7 , 6.0$ improved, are the asymptotic quenched ones for $f_\pi$.

Note also that we convert to MeV by
using the extrapolated value for the $\rho$ mass. The
$f_\pi$ data nicely extrapolate at $132$ MeV thus demonstrating the
consistence between the two scales induced by the $\rho$ mass and
$f_\pi$. (We have to say that since we rely on the perturbative evaluation
of $Z_A$ ~\cite{DC}, it is possible, even if unlikely,
that this good result is
a coincidence. It is clear that a non-perturbative computation of
the multiplicative renormalization constant would be welcome. )

We can get an estimate for $f_\pi$ also from the data in the $\pi$ channel,
modulo a constant. We denote this estimate ($f_{\pi}^{\pi}$), and we quote
its ratio  with the true  $f_\pi$, which is remarkably  stable.

$f_\rho$ is more delicate since  with Wilson action there is
a strong discrepancy between the perturbative and lattice evaluation
for $Z_V$ ~\cite{FR}. Even if in this case the
situation is supposed to be better,
the perturbative results for $Z_V$~\cite{DC}, hence
for the lattice estimate of $f_\rho$, should be
considered with some extra care.

\begin{table}
 \begin {tabular} {||l|l l l l ||} \hline
\multicolumn{5}{||c||}{$\beta = 5.7$} \\
\hline
 &\multicolumn{4}{| c||}{\em k  } \\
\hline
               & 0.186 & 0.190 & 0.194 & 0.198 \\
\hline

$F_{\pi} (MeV)$
               & 243.5( 55)& 219.3( 54)& 191.5( 56)& 160.1( 69)\\
$\frac {f_\pi}{f_\pi^{(\pi)}}$
               & 0.712( 16)& 0.708( 17)& 0.721( 21)& 0.759( 32)\\
$f_\rho^{-1}$
               & 0.281( 11)& 0.309( 13)& 0.341( 19)& 0.383( 47)\\
\hline
\hline
\multicolumn{5}{||c||}{$\beta = 6.0$} \\
\hline
&\multicolumn{4}{| c||}{\em k  } \\
\hline
               & 0.176 & 0.180 & 0.184 & 0.188 \\
\hline
$F_{\pi} (MeV)$
               & 325.9( 35)& 293.6( 38)& 256.3( 39)& 206.4( 39)\\
$\frac{f_\pi}{f_\pi^{(\pi)}}$
               & 0.669(  7)& 0.614(  7)& 0.561(  8)& 0.521(  9)\\
$f_\rho^{-1}$
               & 0.182(  4)& 0.205(  6)& 0.235(  9)& 0.276( 20)\\
 \hline
 \end{tabular}
\protect\caption{Results for the pion and $\rho$ decay constants.}
\protect\label{Fpirho}
\end{table}

The results for baryons are shown in Figs.6 in the form of Ape
invariant mass plot. The plots we believe are self-explanatory,
and demonstrate the consistency of the results for the improved
action at $\beta = 5.7$ and $\beta = 6.0$ with the ones for the Wilson
action at $\beta = 6.3$ and $\beta = 6.0$, while the Wilson results at
$\beta = 5.7$  exhibit clear scaling violations.

A final comment concerns the Proton-$\Delta$ splitting.
In this case we can compare data from different lattices
by normalizing with the running
value of the $\rho$ mass, like in the Ape plot , thus avoiding the
systematics connected with
the chiral extrapolation. Looking at Fig.~7,
we see  that the $\Delta$ and
the proton masses are clearly well resolved, and the results, in the largish
errors, agree with the ones obtained on the other lattices. (
As discussed in ref.~\cite{W63} the data at $\beta=  6.3$ are possibly
biased from the opposite parity partner.)
All the other data are
interpolating almost linearly between the $0$ and the infinite mass limit.
\vskip 1.8truecm

{\bf Acknowledgements}

The numerical simulations were performed  using the Ape computers
in Roma and Pisa. The updating code, the bulk
of the inversion  and the  analysis software were the same
used in refs.~\cite{W63},~\cite{W57},
{}~\cite{W60}: we are especially grateful to Enzo Marinari
for this crucial help.

We owe special thanks to Carlotta Pittori for innumerable  valuable
discussions about improved actions and renormalization constants.
Finally, we wish also thank Herbert Hamber, Enzo Marinari,
Guido Martinelli, Federico Rapuano,
Gian Carlo Rossi, Gaetano Salina and Raffaele Tripiccione
for interesting conversations and useful comments.

The work of MPL is supported by the National Science Foundation,
NSF-PHY 92-00148.

\vfill

\newpage

{\bf Figure Captions.}

\begin{description}
\item[1]
Effective proton mass estimator as a function of the time distance
at $\beta= 5.7$ for the different quark masses.
$24^3\times 32$, squares. $9^3\times 18$, circles.

\item[2]
a) Pion Green functions for the four different quark masses
at $\beta= 6.0$.  The results of
two (one) particle fits are shown as dashed (dotted) lines.
b) as a), for the proton.

\item[3]
$\pi$ (squares, dashed lines) and $\tilde\pi$ (diamonds, dotted lines)
chiral extrapolations at $\beta = 5.7$ (left) and
$\beta = 6.0$ (right). The lines are the results of the fits.

\item[4]
a) The splitting $\rho^2 - \pi^2$
plotted against $\pi^2$.  b) as a), with included the Wilson results.
Everything in unit of the extrapolated $\rho$ mass.

\item[5] a) $f_\pi$ in MeV  versus $\pi^2$ in unit of the extrapolated
$\rho$ mass. b) as a), with included the Wilson results.

\item[6]
a) Ape plot : $m_p/m_\rho$ vs $m_\pi^2/m_\rho^2$ b) as a),
with Wilson results.

\item[7]
$(m_\Delta - m_p)/m_\rho$ vs $m_\pi^2/m_\rho^2$ for the improved and
standard Wilson action.

\end{description}

\end{document}